# Reflecting on Recurring Failures in IoT Development

Dharun Anandayuvaraj
Purdue University, IN, USA
dananday@purdue.edu

James C. Davis
Purdue University, IN, USA
davisjam@purdue.edu

## ABSTRACT

As IoT systems are given more responsibility and autonomy, they offer greater benefits, but also carry greater risks. We believe this trend invigorates an old challenge of software engineering: how to develop high-risk software-intensive systems safely and securely under market pressures? As a first step, we conducted a systematic analysis of recent IoT failures to identify engineering challenges. We collected and analyzed 22 news reports and studied the sources, impacts, and repair strategies of failures in IoT systems. We observed failure trends both within and across application domains. We also observed that failure themes have persisted over time. To alleviate these trends, we outline a research agenda toward a Failure-Aware Software Development Life Cycle for IoT development. We propose an encyclopedia of failures and an empirical basis for system postmortems, complemented by appropriate automated tools.

## CCS CONCEPTS

• **Computer systems organization** → **Cyber-physical systems**.

## KEYWORDS

Internet of Things, Cyber-Physical Systems, Embedded Systems, Safety-Critical Software, Software Engineering, Failure Analysis

## 1 INTRODUCTION

Internet of Things (IoT) realizes a modern world of smart devices interconnected by complex networks. Many IoT systems are software-intensive with hardware components off-the-shelf. IoT systems enable software to directly interact with the physical environment [20]. These interactions are divided into observations of the physical world using *sensors*, and effects on the physical world using *actuators* [20]. A plethora of technological breakthroughs, in batteries, hardware, wireless networking, mobile computing, Cloud services, and machine learning, has enabled widespread adoption of IoT systems [26]. These trends have enabled IoT systems to become pervasive [41] and increasingly interactive with the physical world. Their failures are often safety-critical [27]. Given the complexities of IoT systems, their diverse characteristics enable diverse failures.

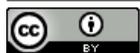

In this paper, we studied IoT failures in the wild to identify engineering challenges. We conducted the first systematic study of IoT failures as reported in the media. Specifically, we studied the sources, impacts, and repair recommendations of IoT failures. We observed trends in failure both within and across application domains. We also observed that failure themes have persisted over time. To alleviate these trends, we outline a research agenda towards a Failure-Aware Software Development Life Cycle.

## 2 BACKGROUND AND RELATED WORK

The study of engineering failures is essential to successful design [32]. In the current state of software engineering research there is a gap to study and learn from IoT *failures*. Researchers have already studied *faults* in IoT systems (*e.g.,* bug studies [19, 28, 40, 44]). The distinction is that a fault is a defect within a system, whereas a failure occurs when one or more faults cause the system to fail in its required function [37]. Not all faults are created equal; the study of failures informs failure mode analysis [22, 34, 36] and helps researchers and engineers prioritize risks [17].

While the software engineering field has benefited from empirical and qualitative failure analysis research [13, 27], no such systematic studies have been conducted to learn from IoT failures. Such works require detailed information about the failures, which are often obtained through sources such as lawsuits and government investigations [27]. These sources of information are still limited for the emerging IoT field. Lacking detailed failure reports, we begin with news reports. We acknowledge that news reports are not ideal data sources, but in the absence of detailed failure reports, it is a starting point. In civil engineering research, news reports have been used as primary data sources to attain information for infrastructure failure interdependencies [43]. This methodology is also used in economics [23], public health research [18], and agriculture [16].

## 3 METHODOLOGY

Our goal is to systematically identify IoT failure events, their sources, and their impacts. Our approach is to conduct a qualitative analysis of failure reports covered by news reports. Specifically, we investigate two research questions:

**RQ1:** What are the common sources of IoT failures?
**RQ2:** What are the common impacts of IoT failures?

### 3.1 Data Collection

We collected IoT failures covered by reputable news sources. We selected *The New York Times* for its focus on general audience [5], and *WIRED* for its popular coverage of technology [10]. We used Google News to search both sources to maintain reproducibility. We used IoT related search terms [1] and limited the search from

---
[1]Search Terms: "iot, internet of thing, cyber physical system, autonomous, cyber security fail, monitor fail, device fail, software fail, computer fail"

January 2015 to October 2021 [2]. Through this search criteria we collected a sample of IoT failures.

For each search, we filtered out articles based on their titles, and subsequently their content. We included any article that described the *failure* of an *IoT system*, following the definitions given in sections 1 and 2. When multiple articles described the same event, the article with greater detail was selected.

### 3.2 Data Analysis

*Content extraction:* As a first-pass analysis we read each article, and identified the sources of failure and the impacts of failure of the IoT systems. If available in the text, repair recommendations were also identified. Sources referenced in the articles were also reviewed for supplementary understanding of the contents.

*Organization of knowledge:* We then performed a structured analysis of the qualitative data to code and categorize the data by sources of faults and taxonomy of faults. We followed the framework of Melo & Aquino [30], which provides definitions and characterizations for studying IoT failures. The framework divides IoT systems into four levels (*i.e.,* layers) in which failures can occur: perception (components for obtaining and processing data), communication (wired and wireless media), service (*e.g.,* data storing and processing services), and application (*e.g.,* user interface). The framework also taxonomizes the behavior and impacts of a failure: duration (transient, permanent, intermittent), fault origin location (internal: within system, external: from environment), semantics (crash, omission, timing, value, arbitrary), change in behavior (soft: altered, hard: inactive), and dimension (software, hardware).

From each news source, we identified 1–3 faults leading to failures. We report on one primary fault for each failure.

## 4 RESULTS AND ANALYSIS

The news articles that matched our search criteria are summarized in Table 1. Our search query yielded 570 results, from which we identified 22 articles that reported on IoT failures. Searching *The New York Times* yielded 14 articles, and *WIRED* yielded 8 articles. The 22 distinct failure events originated from 18 distinct organizations. The events represent 5 categories of IoT applications [8]. The automotive category was the most common (8/22), followed by critical infrastructure (6/22), consumer products (4/22), healthcare (2/22), and aerospace (2/22).

**RQ1: Common sources of IoT failures.** The sources of faults are presented in Figure 1. The most common source of faults originated at the application level (9/22), followed by connectivity at the communication level (7/22). Other sources of faults originated from embedded software and transducers at the perception level, links at the communication level, and the service level. Furthermore, Figure 2 shows that failure-triggering events occurred both within and outside of the system, and that *all faults were traceable to the behavior of software components.*

We observed failure trends both within and across application domains. In the automotive domain, *functional failures* were more common, because of a reliance on cutting-edge (and faulty) computer vision components (ID 7, 8, 10, 11, 12). In the healthcare domain, *the lack of a safe state* led to failures, specifically when network connectivity was lost (ID 19, 20). Across domains, *using a system outside of its intended specification* led to failures in autonomous cars (ID 7, 8, 10, 11, 12), consumer health monitors (ID 19, 20), and smart home products (ID 16). Another cross-domain cause of failure was *insecure remote access and authentication*, affecting critical infrastructure (ID 1, 3, 4, 5, 6), connected cars (ID 9, 14), and consumer products (ID 15, 18). These cybersecurity failures could often be attributed to misplaced trust in vendored components (*i.e.,* the software supply chain). *Improper isolation between safety-relevant vs. application software* led to failures in critical infrastructures (ID 2, 5) and connected cars (ID 9, 13, 14). *Incorrect software evolution* led to failures in a smart home product (ID 22) and an aircraft (ID 17). Additionally, we observed that 14 of the failure events were a result of multiple sources of failures (ID 2, 3, 4, 5, 6, 7, 8, 9, 10, 11, 12, 15, 16, 22). This observation indicates opportunities to better exercise accident management techniques (*i.e.,* Swiss cheese model [33]) for IoT development.

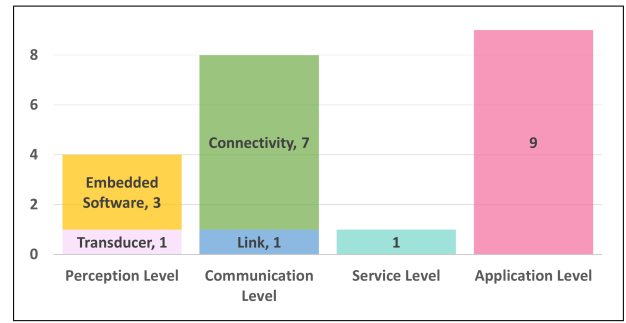

**Figure 1: Sources of faults (*N=22*). Each failure is categorized into one source of fault at the system level.**

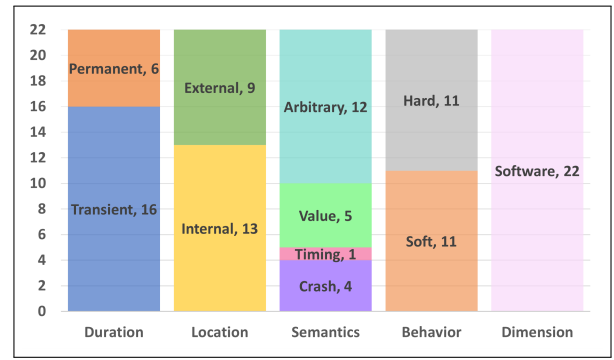

**Figure 2: Taxonomy of faults (*N=22*). Each bar captures a different attribute of each failure.**

**RQ2: Common impacts of IoT failures.** The taxonomy of faults is presented in Figure 2. Many of the systems behaved arbitrarily (12/22) when a fault was active, primarily due to security breaches. Most faults were temporary (16/22), some permanent (6/22), none intermittent. Furthermore, half of the systems exhibited an altered behavior, and half of the systems exhibited an inactive behavior when faults occurred.

General trends amongst the impacts of failures were primarily observed within the application domains (Table 1). Failures in critical infrastructures (ID 1, 2, 3, 4, 5, 6) led to unauthorized access to critical functions, and their impacts were extensive with respect to human impact and monetary cost. Failures in autonomous cars (ID 7, 8, 10, 11, 12) led to fatal collisions, and failures in connected cars (ID 9, 13, 14) led to unauthorized access to safety-critical functions of the car. Failures in consumer products resulted in varying impacts including DDoS attacks (ID 18) and domestic abuse (ID 16). Failures in consumer health monitors (ID 19, 20) led to a false sense of safety. Failures in aerospace systems led to severe impacts such as ill-timed state changes (ID 21) and aircraft crashes (ID 22).

---

[2]Google News Search Syntax: "[SearchTerm] site:[SourceWebsite] after:[StartDate: Year-Month-Day] before:[EndDate: Year-Month-Day]"

Table 1: IoT failures, organized by Cat(egory). *'ID' denotes row in artifact (§8). Repair recommendations marked ' + ' are directly from articles.*

| Cat. | ID | Date | System | Failure Source | Failure Impact | Repair Recommendation |
|---|---|---|---|---|---|---|
| Critical infrastructure | 1 | 2021 | Water works | Lack of seperation between IT & OT network | City's water poisoned | Remove remote access[+] |
| | 2 | 2021 | Oil pipeline | Hack of business network; lack of separation between business & operation networks | Shutdown of a major oil pipeline | Isolate critical infrastructure network[+] |
| | 3 | 2021 | Public transportation | Zero day attack in remote access software; malware web shells | Unauthorized access; $370,000 response cost | Scan periodically for backdoor web shells[+] |
| | 4 | 2019 | Water treatment facility | Un-updated discharged personnel authentication; unsecure remote access | Unauthorized access; altered cleaning & disinfecting process | |
| | 5 | 2015 | Power plant | Unsecure supply chain network; malware repositories | Unauthorized access; leak of critical information | |
| | 6 | 2015 | Power distribution center | Unsecure remote access; lack of 2FA for SCADA | Loss of power for 230,000 residents | Enable 2FA for safety-critical remote access[+] |
| Automotive | 7 | 2019 | Autonomous car | Failed to detect parked vehicle, stop sign, flashing red lights; ineffective driver engagement monitor | Fatal collision | Stringent user engagement monitor[+] |
| | 8 | 2019 | Autonomous car | Failed to detect merging vehicle; neglected radar data; ineffective driver engagement monitor | Fatal collision | Create redundancy in object detection system[+]; stringent user engagement monitor[+] |
| | 9 | 2019 | Connected car | Default authentication for user accounts; lack of isolation for safety-critical functions | Unauthorized access to safety-critical functions | Disable default authentication[+] |
| | 10 | 2018 | Autonomous car | Ineffective parked vehicle identification; ineffective driver engagement monitor | Fatal collision | Stringent user engagement monitor[+] |
| | 11 | 2018 | Autonomous car | Failed to detect road barrier; ineffective driver engagement monitor | Fatal collision | Stringent user engagement monitor[+] |
| | 12 | 2016 | Autonomous car | Failed to detect turning vehicle & stop at collision; ineffective driver engagement monitor | Fatal collision | |
| | 13 | 2015 | Connected car | Zero day exploit of entertainment system firmware | Unauthorized access to safety-critical functions | Isolate safety-critical functions[+]; auto monitor CAN bus[+] |
| | 14 | 2015 | Connected car | Unsecure remote access protocol (SMS, SSH w/ default key) in telematics dongle | Unauthorized access to safety-critical functions | Secure remote access[+]; isolate safety-critical functions[+] |
| Consumer products | 15 | 2019 | Connected real-time OS | Networking protocol bug in COTS; Lack of software bill of materials | Exploitable vulnerabilities | Maintain software bill of materials[+] |
| | 16 | 2018 | Smart home products | Lack of all users' consent; abuse enabling user experience | Domestic abuse; psychological stress | |
| | 17 | 2016 | Smart thermostat | Bug in software update | Battery depletion; Unprompted temperature decrease | |
| | 18 | 2016 | Old connected devices | Processors with default authentication | Distributed denial-of-service (DDoS) botnet; internet outage | Disable default authentication[+] |
| Healthcare | 19 | 2019 | Smart baby vital monitor | Device to server to phone app connection loss | False sense of safety | |
| | 20 | 2019 | Smart diabetes monitor | Device to server to phone app connection loss | False sense of safety | Alert when connection lost[+] |
| Aerospace | 21 | 2019 | Spacecraft | Vendored software with bug caused early communication query | Early state change depleted fuel | |
| | 22 | 2019 | Aircraft | Faulty sensor; lack of sensor redundancy; lack of updated system training | Catastrophic crashes | Create redundancy for critical systems[+] |

## 5 DISCUSSION

Since the unique challenges of IoT lie in the complexity of system design [9], it is beneficial to study failures from the system perspective. Previous works have largely examined *faults* present in IoT *software* rather than *failures* of IoT *systems* [19, 28, 40, 44]. Because of the focus on faults, they contain limited information about failure impacts (risks). Moreover, these works primarily study open-source software, which may not be representative of commercial systems [38]. In contrast, our data provides system-level information about the failures of IoT systems. The use of news reports documenting IoT failures enabled us to identify distinct faults along with their human impacts. This approach enables engineers to focus on faults that lead to catastrophic failures in IoT systems.

Our results indicate that half of the failures were due to cybersecurity issues (ID 1, 2, 3, 4, 5, 6, 9, 13, 14, 15, 18). This finding adds weight to government and industry calls to reinforce the security of IoT systems. Researchers could investigate whether these failures could have been prevented by following government guidelines [4], industry guidelines [6], and international standards [24].

Lastly, we observed echoes of past failures in modern IoT systems. For example, software evolution was a cause of the 1980s Therac-25 failure [27], and is still manifesting (ID 22). Within critical infrastructure, lack of isolation of safety-critical functions and insecure remote access led to unauthenticated access (ID 2, 5), a concern also raised by the National Research Council in 2007 based on failures in the 1990s-2000s [13].

## 6 RESEARCH AGENDA

Although we took a system-level perspective on IoT failures, it appears that many of these failures can be traced to problems in software and system engineering (Figure 2). The persistence of failure themes over time is unsurprising — these are some of the core challenges of software engineering — but we suggest that the software engineering discipline might benefit from development processes that place a greater emphasis on mitigating past failures. We therefore propose three research directions towards a Failure-Aware Software Development Life Cycle for IoT (Figure 3).

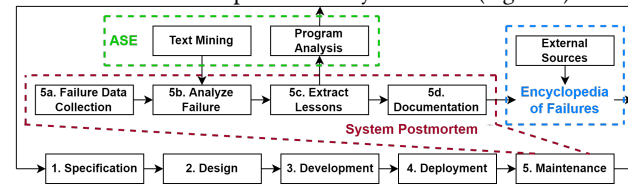

**Figure 3: A Failure-Aware Software Development Life Cycle.**

*Infrastructure: A Failure Encyclopedia.* To help IoT engineers anticipate failures, we believe they would benefit from an encyclopedia of previous IoT failures modeled on Table 1. The closest existing knowledge base is the CVE database of cybersecurity vulnerabilities [2], which deliberately omits root cause information — protecting affected users, but limiting the value for other engineering teams. A catalog of case studies could outline the failure, the underlying fault(s), the impact, and lessons learned from a system failure. Case studies could be built from within teams and organizations, as well as from external sources such as news reports or other organizations. These case studies could inform software engineering judgment [29, 37]. Additionally, they could help engineers determine the extent to which failures can be attributed to different stages of the engineering process. This would provide engineers with a heuristic for investing their time in different stages.

*Process: An Empirical Basis for Postmortems.* Our analysis of failures was restricted because we could only observe what was reported by journalists. IoT engineers working on the affected products could conduct a more detailed analysis to benefit their own and other teams, through a failure postmortem. Although postmortems are widely recommended [11, 12, 14], they are often omitted [15, 25, 39]. We know surprisingly little about postmortem practices in software engineering. We therefore recommend research to establish an empirical basis for software failure postmortems. For example, what are effective personal and team practices to collect, analyze, and document system failures? How can postmortem knowledge be integrated into the software development cycles, managing the tradeoff between agility and risk management? What mechanisms would support measuring the impact of postmortems in mitigating or recovering from failures?

*Tools: Automation.* There are many opportunities to automate elements of this research agenda. We outline one three-part sequence based on Figure 3. First, there are many computing failures in the news, including both IoT and IT software. Researchers could leverage text mining techniques (NLP) to extract system postmortem information from diverse representations, including news reports [42], user complaints [21], and open-source issue reports. This would facilitate the organization of a large encyclopedia of failures. Then, an engineering team would want to query this database for relevant failures to guide their system design or maintenance work; how can they filter thousands of cases down to the ones relevant in their context, or use machine learning to transfer lessons across contexts [35]? Finally, during validation, an engineering team might want to scan their system model or their codebase for known hazards, *e.g.,* using program analysis techniques [1, 3, 7] to identify anti-patterns. In the IoT context, an end-to-end scan might be difficult, suggesting a human-in-the-loop approach.

## 7 THREATS TO VALIDITY

We note two sources of methodological bias. First, the raw data: journalists might dramatize the impacts of IoT failures, and editors may sway which events are covered [31]. Second, only one researcher coded the data; a second author reviewed the process. News articles are designed for lay readers; coding was straightforward.

## 8 CONCLUSION

In this work, we studied real-world IoT systems to better understand the sources and impacts of IoT failures. We observed persistent failure trends both within and across application domains. We outlined a research agenda towards a Failure-Aware Software Development Life Cycle for IoT development.

**Data:** https://doi.org/10.5281/zenodo.7033016.